# On Validation of Search & Retrieval of Tissue Images in Digital Pathology

H.R. Tizhoosh

*Kimia Lab, Mayo Clinic, Rochester, MN, USA*

**Abstract** - Medical images play a crucial role in modern healthcare by providing vital information for diagnosis, treatment planning, and disease monitoring. Fields such as radiology and pathology rely heavily on accurate image interpretation, with radiologists examining X-rays, CT scans, and MRIs to diagnose conditions from fractures to cancer, while pathologists use microscopy and digital images to detect cellular abnormalities for diagnosing cancers and infections. The technological advancements have exponentially increased the volume and complexity of medical images, necessitating efficient tools for management and retrieval. Content-Based Image Retrieval (CBIR) systems address this need by searching and retrieving images based on visual content, enhancing diagnostic accuracy by allowing clinicians to find similar cases and compare pathological patterns. Comprehensive validation of image search engines in medical applications involves evaluating performance metrics like accuracy, indexing, and search times, and storage overhead, ensuring reliable and efficient retrieval of accurate results, as demonstrated by recent validations in histopathology.

## Introduction

Images play a critical role in modern medicine, providing essential information that aids in diagnosis, treatment planning, and monitoring of diseases [1]. Fields like radiology and pathology heavily rely on the accurate interpretation of images [2]. Radiologists analyze medical imaging such as X-rays, CT scans, and MRIs to diagnose conditions ranging from broken bones to complex diseases like cancer [3]. Pathologists examine tissue samples through microscopy to detect abnormalities at a cellular level, which is crucial for diagnosing cancers, infections, and other diseases [4]. The volume and complexity of medical images have grown exponentially with technological advancements, necessitating efficient tools for their management and retrieval [5].

Content-Based Image Retrieval (CBIR) [6] systems are designed to search and retrieve images from small and large databases based on the visual content within the images rather than metadata or keywords. In medicine, CBIR can enhance diagnostic accuracy by enabling clinicians to find visually similar cases, compare pathological patterns, and draw insights from large image datasets [7,8]. This is particularly valuable in situations where unique or rare conditions are involved, providing access to similar cases that might not be documented in textual records.

# Comprehensive Validation of Image Search Engines

To ensure the reliability and effectiveness of image search engines in medical applications, comprehensive validation is necessary [9]. Validation involves evaluating several performance metrics that reflect the system's ability to retrieve accurate results quickly and efficiently. The following criteria are essential for a thorough assessment. Lahr et al. [10] recently validated image retrieval in histopathology and provided a comprehensive framework for testing. They not only looked at accuracy measurements via F1-score but also quantified indexing and search times, measured overhead storage for indexing and counted failures to retrieve.

## 1. F1-Score of Top 1 Search Result

The F1-score is a measure of a test's accuracy, considering both precision (the number of true positive results divided by the number of all positive results, including those not identified correctly) and recall (the number of true positive results divided by the number of results that should have been identified). The F1-score of the top 1 search result indicates how often the very first image returned by the search engine is both relevant and accurate. This metric is critical in medical applications where the first match often carries significant weight in diagnostic decision-making. However, to avoid reducing search to mere classification, the evaluation must consider top-1 accuracy alongside top-3 and top-5 retrievals.

## 2. F1-Score of Majority Vote among Top 3 Search Results

Evaluating the F1-score based on the majority vote among the top 3 search results can provide a more robust assessment of the search engine's accuracy [11]. This metric considers the top three returned images and determines if the majority are relevant. This approach mitigates the risk of occasional errors and gives a clearer picture of the search engine's performance in providing reliable results.

## 3. F1-Score of Majority Vote among Top 5 Search Results

Similar to the majority vote among the top 3 results, the F1-score for the top 5 search results further enhances the robustness of the evaluation. Including more images in the majority vote increases the likelihood that the retrieved set contains the correct or most relevant images, providing a more comprehensive assessment of the search engine's ability to retrieve accurate results consistently. Going above five retrievals will burden the pathologist if visual inspection and verification become necessary.

The example in Figure 1 demonstrates the necessity for considering F1-score of top-1, majority of top-3, and majority of top-5.

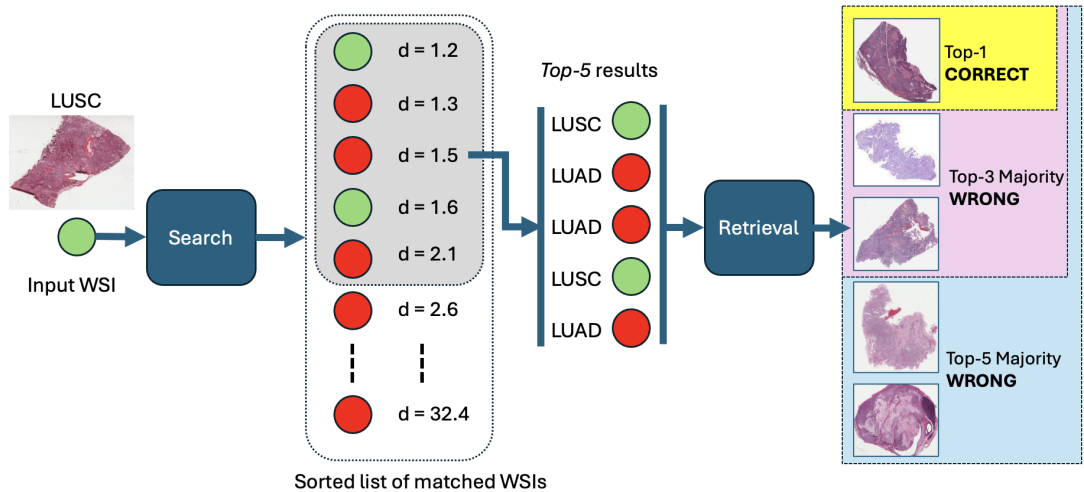

*Figure 1. An example of lung squamous carcinoma (LUSC) and lung adenocarcinoma (LUAD) demonstrated the significance of looking at top-1, top-3 and top-5 retrievals simultaneously.*

## 4. Speed of Indexing and Search

The efficiency of an image search engine is not only about accuracy but also speed. In medical environments, timely access to information can be critical. The speed of indexing (how quickly new images can be added to the database) and search (how quickly the engine can retrieve relevant images) are vital metrics. A system that indexes and retrieves images rapidly can significantly enhance clinical workflows and decision-making processes.

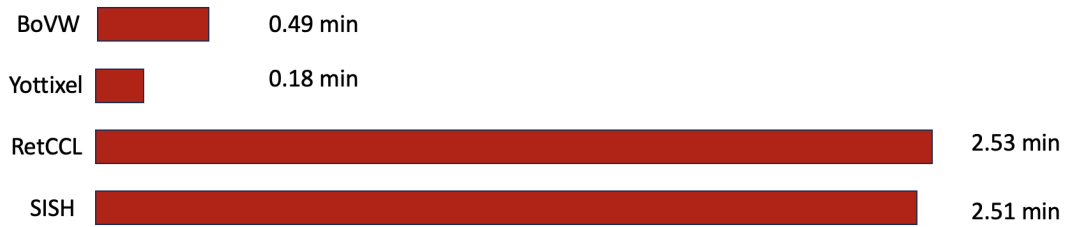

*Figure 2. Indexing and searching times may be platform-independent but provide direct useful information to determine the value of image search engine (see [10]).*

Explicit time measurements are often more suited than theoretical upper bounds, such as Big O notation, for assessing the practical limitations of image retrieval systems (see Figure 2). While "Big O notation" provides a high-level understanding of the algorithm's efficiency by describing its behavior as the input size grows, it doesn't account for the actual runtime performance on real-world data or the effects of hardware, software, and implementation details. Explicit time measurements, on the other hand, offer concrete performance data under specific conditions, allowing for a more accurate evaluation of how the system performs in practice. This practical insight is crucial for optimizing and fine-tuning image retrieval systems to meet real-world demands and constraints.

An important aspect of time measurement is to consider the time of "indexing and search" as the total search time. Since we cannot conduct a search without first indexing the input image, it follows that the search time includes both the indexing time and the actual search time.

## 5. Storage Overhead Requirement for Indexing

Given the high cost of storage solutions like solid-state drives (SSDs), the storage overhead required for indexing each image is an important consideration [12]. Efficient use of storage without compromising the accuracy and speed of retrieval is crucial for the practical deployment of CBIR systems in medical settings. Systems that require excessive storage can be prohibitively expensive and impractical for large-scale implementation (see Figure 3).

The democratization of AI is obviously a paramount factor. Considering the gigapixel nature of WSIs, many small clinics and community hospitals may face difficulties on top of an "already financially and operationally stressed healthcare system" [13]. Most labs and hospitals in developing countries do not have the financial resources to purchase high-performance SSD storage to archive WSIs [14]. In light of these limitations, asking for excessive extra storage to run a search engine appears illusory, a challenge that is widely recognized by many stakeholders [15].

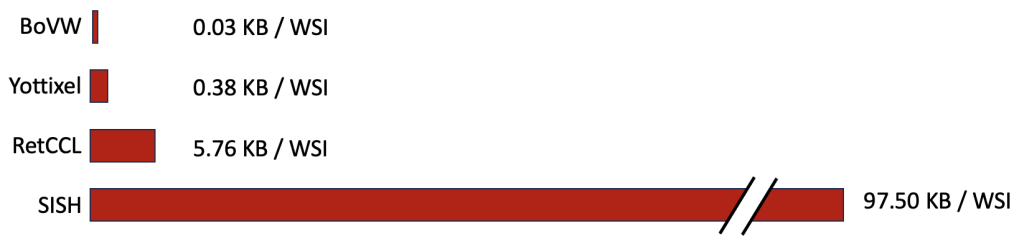

Figure 3. Some search engines are impractical due to excessive need to store indexing information per whole slide image [10].

## 6. Failures to Retrieve

Finally, the reliability of the search engine is measured by the number of times the search fails to return a relevant result. Search failures can have serious implications in medicine, potentially leading to misdiagnosis or delayed treatment. A robust CBIR system should have a minimal failure rate, ensuring that relevant images are almost always retrieved.

## 7. Ranking Search Methods

To present all these evaluation metrics in a simple manner to help the user (researcher, physician) to select the best method, Lahr et al. ranked all search engines for each of the above-mentioned categories based on their performance and calculated an overall performance ranking (Figure 4). Such performance rankings are urgently needed to establish a realistic picture of image search engines in histopathology.

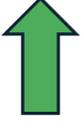

*Figure 4. Lahr et al. [10] validated four image search engines (Yottixel, SISH, BoVW and RetCCL) in multiple configurations for accuracy (top-1, majority vote at top-3, majority vote at top-5), speed (indexing time, indexing+searching time), robustness (number of times retrieval failed), and indexing storage (total overhead). The overall rank R was then reported as the overall performance.*

## 8. Still Missing: Penalizing Violators of Occam's Razor

The only quantitative measure missing in Lahr et al.'s validation appears to be the analysis of model size and its maintainability. This omission is noteworthy given the increasing importance of model scalability and sustainability. It is possible that the authors did not consider this aspect because the search engines they validated were using "ordinary" deep models, rather than the large models that are now being touted as foundational.

The size of a model is a critical factor that impacts several dimensions of its practical application. Larger models, while potentially offering improved performance, come with significant drawbacks. One of the primary concerns is the increased carbon footprint associated with training and deploying these models. Larger models require more computational resources, leading to higher energy consumption and environmental impact. This aspect is particularly important in the context of growing awareness and responsibility towards sustainable AI practices. As well, larger models introduce greater complexity, which can complicate their integration and operation within existing systems. The maintainability of large models is a significant concern. As models scale up, the effort required to update, fine-tune, and adapt them to new data or evolving requirements increases substantially. This can lead to a new type of disparity in healthcare as small clinics and hospitals often do not have necessary resources to pay for maintenance.

## Conclusion

In the realm of medical imaging, the validation of image search engines must be rigorous and comprehensive, incorporating multiple performance metrics to ensure reliability and effectiveness. Evaluating the F1-scores of top search results, the speed of indexing and search, storage overhead requirements, and the rate of search failures provides a holistic view of the system's performance. As well, the model size and complexity must be evaluated to support democratization of AI and avoid disparities in healthcare delivery. Only validation reports that consider all these aspects can be trusted to accurately reflect the capabilities of CBIR systems in medicine. Ensuring these systems are robust, efficient, and reliable is essential for supporting clinicians in making accurate and timely diagnosis, ultimately improving patient outcomes.

As an ethical note, known accuracy, failure, storage and robustness issues with methods must be reported. Omitting such results constitutes academic dishonesty.